\documentclass{emulateapj}
\usepackage{apjfonts}

\newcommand{\gapprox}{\mathrel{\mathpalette\@versim>}}
\newcommand{\lapprox}{\mathrel{\mathpalette\@versim<}}
\newcommand{\propapprox}{\mathrel{\mathpalette\@versim\propto}}

\shorttitle{Expansion of N103B} 
\shortauthors{WILLIAMS ET AL.}

\begin{document}

\title{The Expansion of the Young Supernova Remnant 0509-68.7 (N103B)}

\author{Brian J. Williams,\altaffilmark{1}
William P. Blair,\altaffilmark{2}
Kazimierz J. Borkowski,\altaffilmark{3}
Parviz Ghavamian,\altaffilmark{4}
Sean P. Hendrick,\altaffilmark{5}
Knox S. Long,\altaffilmark{6}
Robert Petre,\altaffilmark{1}
John C. Raymond,\altaffilmark{7}
Armin Rest,\altaffilmark{6}
Stephen P. Reynolds,\altaffilmark{3}
Ravi Sankrit,\altaffilmark{6}
Ivo R. Seitenzahl,\altaffilmark{8}
P. Frank Winkler,\altaffilmark{9}
}

\altaffiltext{1}{NASA Goddard Space Flight Center, Greenbelt, MD 20771; brian.j.williams@nasa.gov}
\altaffiltext{2}{Johns Hopkins University, Baltimore, MD}
\altaffiltext{3}{North Carolina State University, Raleigh, NC}
\altaffiltext{4}{Towson University, Towson, MD}
\altaffiltext{5}{Millersville University, Millersville, PA}
\altaffiltext{6}{Space Telescope Science Institute, Baltimore, MD}
\altaffiltext{7}{Harvard-Smithsonian Center for Astrophysics, Cambridge, MA}
\altaffiltext{8}{University of New South Wales, Australian Defence Force Academy, Canberra, ACT 2600, Australia}
\altaffiltext{9}{Middlebury College, Middlebury, VT}

\begin{abstract}

We present a second epoch of {\it Chandra} observations of the Type Ia LMC SNR 0509-68.7 (N103B) obtained in 2017. When combined with the earlier observations from 1999, we have a 17.4-year baseline with which we can search for evidence of the remnant's expansion. Although the lack of strong point source detections makes absolute image alignment at the necessary accuracy impossible, we can measure the change in the diameter and the area of the remnant, and find that it has expanded by an average velocity of 4170 (2860, 5450) km s$^{-1}$. This supports the picture of this being a young remnant; this expansion velocity corresponds to an undecelerated age of 850 yr, making the real age somewhat younger, consistent with results from light echo studies. Previous infrared observations have revealed high densities in the western half of the remnant, likely from circumstellar material, so it is likely that the real expansion velocity is lower on that side of the remnant and higher on the eastern side. A similar scenario is seen in Kepler's SNR. N103B joins the rare class of Magellanic Cloud SNRs with measured proper motions.


\end{abstract}

\section{Introduction}
\label{intro}

The supernova remnant (SNR) 0509-68.7 is one of the most luminous X-ray sources in the Large Magellanic Cloud (LMC), despite being only $\sim 30''$ in diameter (about 3.6 pc at the distance of the LMC).\footnote{This remnant is often labeled as N103B, though this labeling is incorrect. "N103B" refers to a nebula first discovered at optical wavelengths by \citet{henize56}, but that nebula is {\bf not} the SNR (it is an H II region several arcminutes in diameter to the SW of the remnant). SNR 0509-68.7 was first identified in radio waves by \citet{mathewson73}. Nonetheless, the N103B misnomer has stuck, and is more widely associated with this remnant than the original nebula, so we use the name for the remainder of this work.}  {\it ASCA} observations of N103B were presented in \citet{hughes95}, who first identified the remnant as the result of a Type Ia supernova (SN), a conclusion that has been confirmed by several authors \citep{lewis03,badenes09,lopez11,yang13}. A single light echo was detected by \citet{rest05}, who derive an age of 860 yr for the remnant, broadly consistent with its small size (remnants of comparable size in the LMC are $<1000$ yr old). \citet{ghavamian17} used integral field spectroscopy of the Balmer-dominated shocks to detect broad H$\alpha$ emission having a width as high as 2350 km s$^{-1}$. They derive an age of 685 years, assuming the remnant is in the Sedov phase. Recently obtained light echo spectroscopy has shown that the spectrum of the light echo is consistent with an SN Ia origin (A. Rest, in preparation). 

In \citet[][henceforth W14]{williams14}, we used {\it Spitzer} imaging and spectroscopy to show that the remnant appears to be interacting with dense circumstellar material (CSM) ($n_{0} \sim 10$ cm$^{-3}$), remarkably similar to the densities observed in Kepler's SNR by \citet{williams12}. We suggested there that N103B is an LMC older cousin of Kepler's SNR, and thus far, these two remnants are the only two members of the class of Type Ia remnants interacting with dense CSM hundreds of years after the explosion. \citet{li17} present optical imaging and spectroscopy of N103B, concluding that the lack of emission in the eastern half of the remnant is caused by the asymmetric distribution of the CSM due to the high proper motion of the progenitor binary system toward the west.

In this letter, we report a new epoch of X-ray observations with {\it Chandra} from 2017, which we use to measure the expansion of the remnant over a 17.4 year baseline. While X-ray proper motion measurements of Galactic remnants are commonplace, only a few remnants in the Magellanic Clouds have a high enough expansion velocity to allow for an expansion measurement. One example of this is the young Type Ia remnant 0509-67.5, which has had expansion measurements reported in the literature in both optical \citep{hovey15} and X-ray \citep{helder10,roper18} wavelengths. As another example, \citet{xi18} have used {\it Chandra} observations of 1E0102.2-7219 in the Small Magellanic Cloud to observe the proper motions in that remnant.

The paper is organized as follows. In Section~\ref{obs}, we detail the X-ray observations and data reduction, and the attempts to align the two epochs to a common coordinate system. In Section~\ref{disc}, we report the results of our measurements and discuss their interpretation. Section~\ref{conclusions} serves as a summary of our findings. Throughout this paper, we convert our measurements made in image units to the more useful physical quantity of km s$^{-1}$, taking advantage of the known distance to the LMC of 50 kpc \citep{pietrzynski13}.

\begin{figure}[htb]
\begin{center}
\includegraphics[width=7cm]{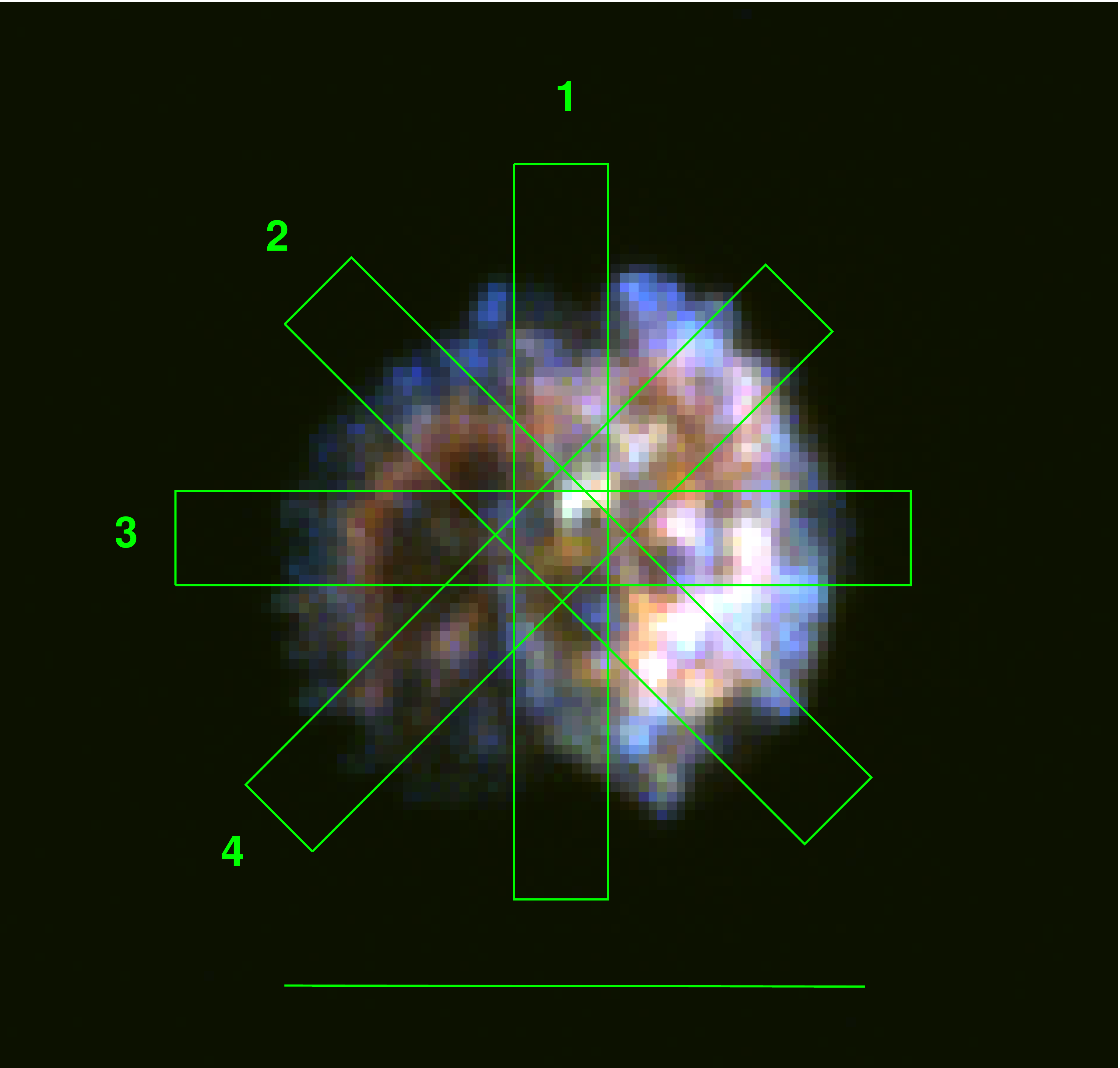}
\caption{A {\it Chandra} X-ray image of N103B, with 0.5$-$1.2 keV in red, $1.2-2.0$ keV in green, and $2.0-7.0$ keV in blue, overlaid with our four profile extraction regions, as described in the text. Each region is 10 pixels wide, where a pixel is the native {\it Chandra} pixel scale of $0.492''$. The scale bar at the bottom is 30$''$ in length.
\label{n103b}
}
\end{center}
\end{figure}

\section{Observations}
\label{obs}

We conducted a new epoch of {\it Chandra} imaging observations of N103B in the spring of 2017, with a total of 400 ks spread over 12 separate observations between Mar 20 and Jun 1. We used the ACIS-S array for these observations, placing the remnant (only $\sim 30''$ in diameter) close to the center of the optical axis of the telescope on the S3 chip, where {\it Chandra's} spatial resolution is best. The 1.7$-$7 keV image from these 2017 observations is shown in Figure~\ref{n103b}. To measure the expansion of the remnant, we compared our 2017 observations with the earliest epoch: a 1999 Dec 4 observation for 40 ks (PI - G. Garmire). Our method for fitting the proper motion involves shifting one epoch with respect to another, accounting for the uncertainties in each epoch. This technique is described more fully in \citet{williams17}  and \citet{katsuda08}, but briefly, we extract brightness profiles from the image in units of counts, with the square root of the number of counts as the uncertainty on each pixel. One epoch, generally the second, is used as the "reference" epoch, with the other epoch shifted until the total $\chi^{2}$ value is minimized. 

The effect we are measuring is quite small (sub-arcsecond, see Section~\ref{disc}), so we took care to minimize or eliminate any potential sources of systematic uncertainty in our measurements. First, we opted not to combine the data from our 12 observations into a single event file, as this could create biases in the resulting FITS files at the sub-pixel level which would be difficult, if not impossible, to quantify. For our "reference" frame, we used the deepest single observation from our 2017 observations: a 60 ks observation (Obs ID 19923) begun on Apr 26. These data, along with the 1999 data (which combine for a time baseline of 17.4 yr), were both processed in identical fashion with the {\it chandra\_repro} script in CIAO version 4.9 (using version 4.7.3 of the {\it CalDB}).

\subsection{Image Registration and Alignment}
\label{alignment}

The standard methods for aligning images in the world coordinate system (WCS) for a proper motion measurement involve either registering point sources detected in the image with known sources from external catalogs or aligning on common point sources within the images from each epoch, allowing for at least a relative alignment. While the former method is obviously preferred, X-ray analysis often relies on the latter, due to the relative paucity of point sources in the X-ray band with known optical counterparts. N103B is an LMC remnant; however, its location (R.A. = $05^{\rm h}08^{\rm m}59^{\rm s}$, Decl. = $-68\degr 43\arcmin 34\arcsec$, J2000.) is well outside the main bar of the LMC and unfortunately, the number of point sources in the field of view is quite low. We restricted ourselves to point sources on the S3 chip (within $4'$ of the remnant), because Chandra's point-spread function (PSF) degrades quickly as a function of off-axis angle. 

To search for point sources in the events files from the two epochs, we first used the CIAO task {\it wavdetect}, as recommended by the {\it Chandra X-ray Center}. This task "found" a few dozen point sources in the image, but most were false positives. A relatively simple search by eye confirmed that only five of these sources were real and detected in both epochs. Using these five sources as input, we created a transformation matrix file using the {\it wcs\_match} task, then used {\it wcs\_update} to align the 1999 epoch 1 image to the 2017 epoch 2 image.

Unfortunately, the results of this alignment were not accurate enough for a robust measurement. When we attempted to measure the proper motion of the leading edge of the emission (presumably the shock front), our results varied substantially, with results approaching 15,000 km s$^{-1}$ in some places and {\em negative} 5000 km s$^{-1}$ in others! We re-did the alignment using another CIAO tool, {\it srcextent}, but the results were similarly wildly varying depending on location within the remnant.

\begin{figure}[htb]
\begin{center}
\includegraphics[width=7cm]{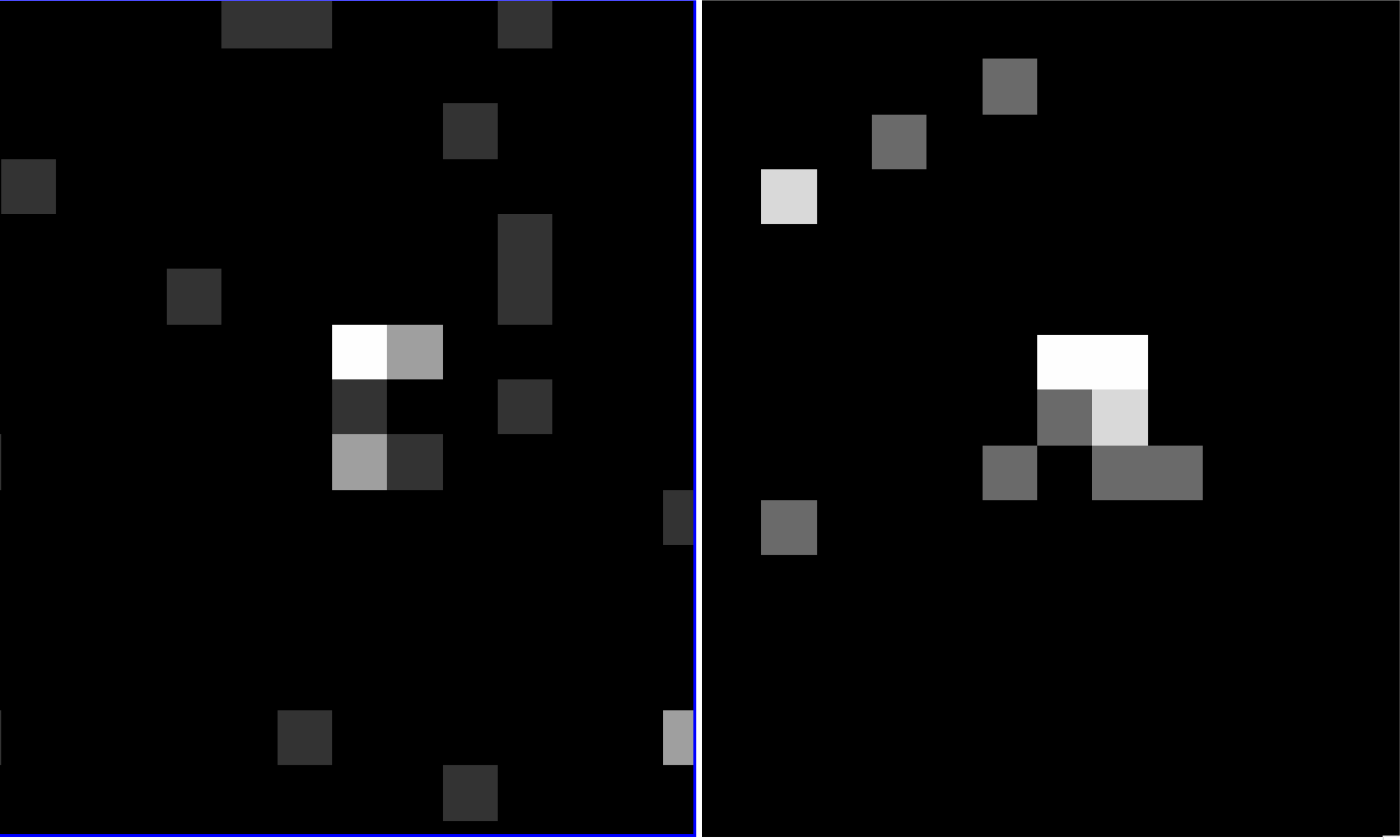}
\caption{One of the sources detected in the field of view, located approximately $1.5'$ from the remnant. The counts image from the 1999 observation is shown on the left, while the 2017 image is shown on the right. Each image is shown at the native {\it Chandra} pixel scale of $0.492''$. The source contains 16 counts in the 1999 data and 25 counts in the 2017 data. 
\label{faintsource}
}
\end{center}
\end{figure}

Upon further inspection, we concluded that most of the point sources (all but one) used for alignment in both the {\it wavdetect} and {\it srcextent} methods do not have a strong enough detection for these algorithms to fit a PSF and determine an accurate location. As an example of what we mean by this, we show one of our sources in Figure~\ref{faintsource}. That this source is real is unquestionable, and it appears in nearly the same location in both epochs. But Figure~\ref{faintsource} shows why any localization algorithms would not be able to report a location to the accuracy that we require.  This source contains only about 15-25 photons total, depending on the epoch, nowhere near enough counts to get a two-dimensional centroid accurate to the sub-arcsecond level. For example, using equation 13 from \citet{kim07}, we find positional uncertainties in the source shown in Figure~\ref{faintsource} of 0.44$''$ and 0.35$''$ in the first and second epoch, respectively. The accuracy of the location is of utmost importance here, since the signal we are searching for is so small. For reference, even with {\it Chandra} and a time baseline of 17.4 years, a 5000 km s$^{-1}$ blast wave at a distance of 50 kpc would move only $\sim 0.37''$ during that time, or about 3/4 of a {\it Chandra} pixel.

We explored other options as well. There are a few knots of emission near the center of the remnant that could, in principle, serve as markers for alignment\footnotemark. However, not only are these sources relatively diffuse (several arcseconds in extent), there is also no way of knowing that these knots do not have their own motion or slightly varying surface brightness profiles over the 17.4 years of evolution of the remnant. We also opted not to use CIAO's sub-pixelization algorithms. There is too much potential for the introduction of a significant systematic uncertainty by reducing the pixel size on the same length scale as the signal we are trying to measure. 

\footnotetext{From the point of view of image transformation, it is a simple matter to simply arbitrarily define any point one wishes as a "fixed source" for alignment between the two epochs.}

Ultimately, the uncertainties involved in obtaining image registration down to a fraction of a pixel led us to focus on measuring one thing that doesn't require knowledge of the WCS registration: the remnant's diameter along various axes. As we show below, we chose to measure the diameter of N103B in four directions (see Figure~\ref{n103b}), effectively forming two orthogonal cardinal coordinate systems. Before making these measurements, we made one final filtering of the data. The ACIS array has suffered significant degradation at low energies due to contaminant buildup since launch. At low energies, the difference between the effective area in 1999 and 2017 is quite significant. Thus, we only considered counts at energies above 1.7 keV (up to 7 keV), high enough to ensure a nearly similar effective area to the 1999 observations while still capturing the strong $\sim 1.8$ keV Si K$\alpha$ line.


\section{Measurements and Discussion}
\label{disc}

\begin{figure}[htb]
\begin{centering}
\includegraphics[width=6cm]{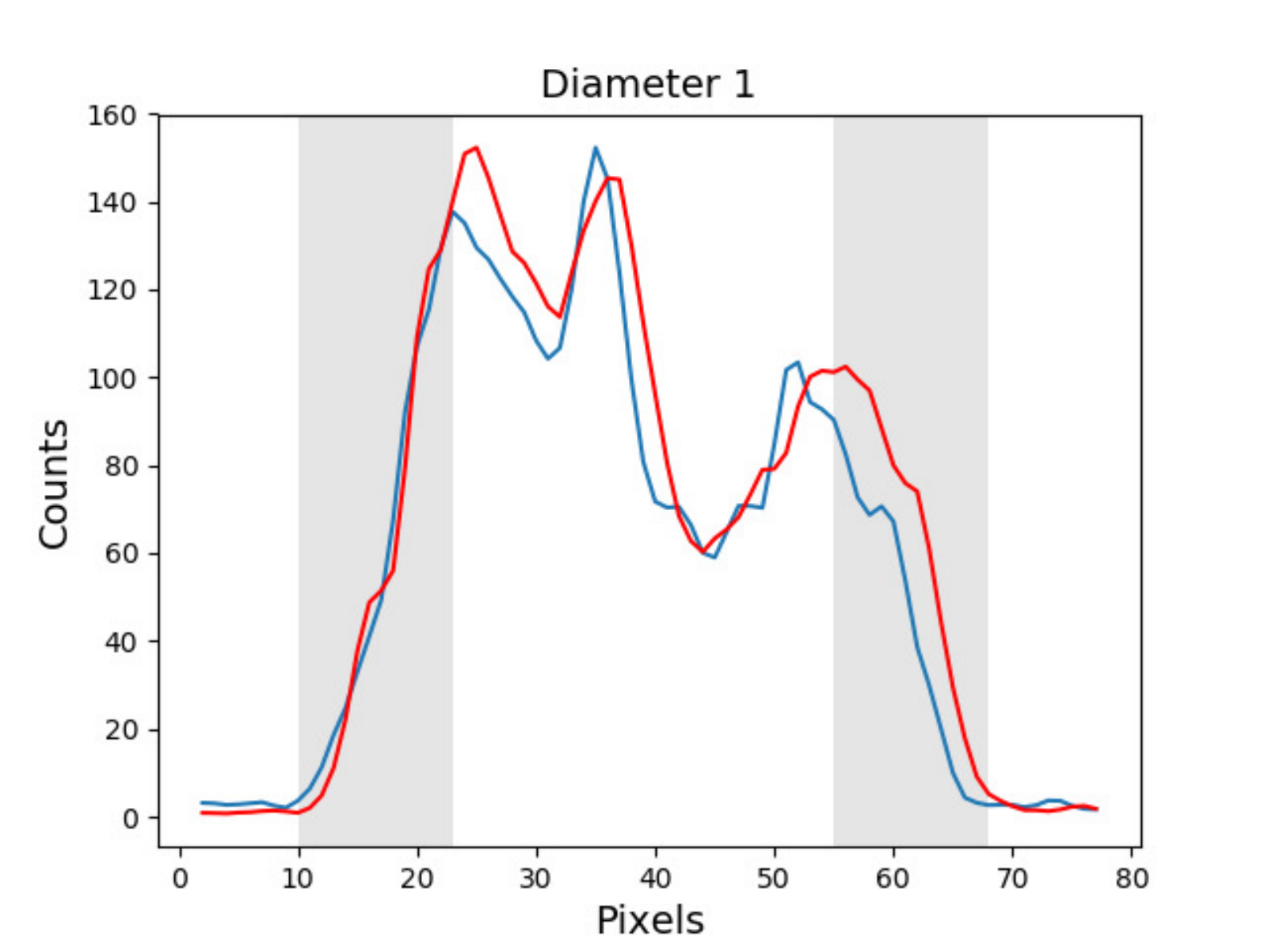}
\includegraphics[width=6cm]{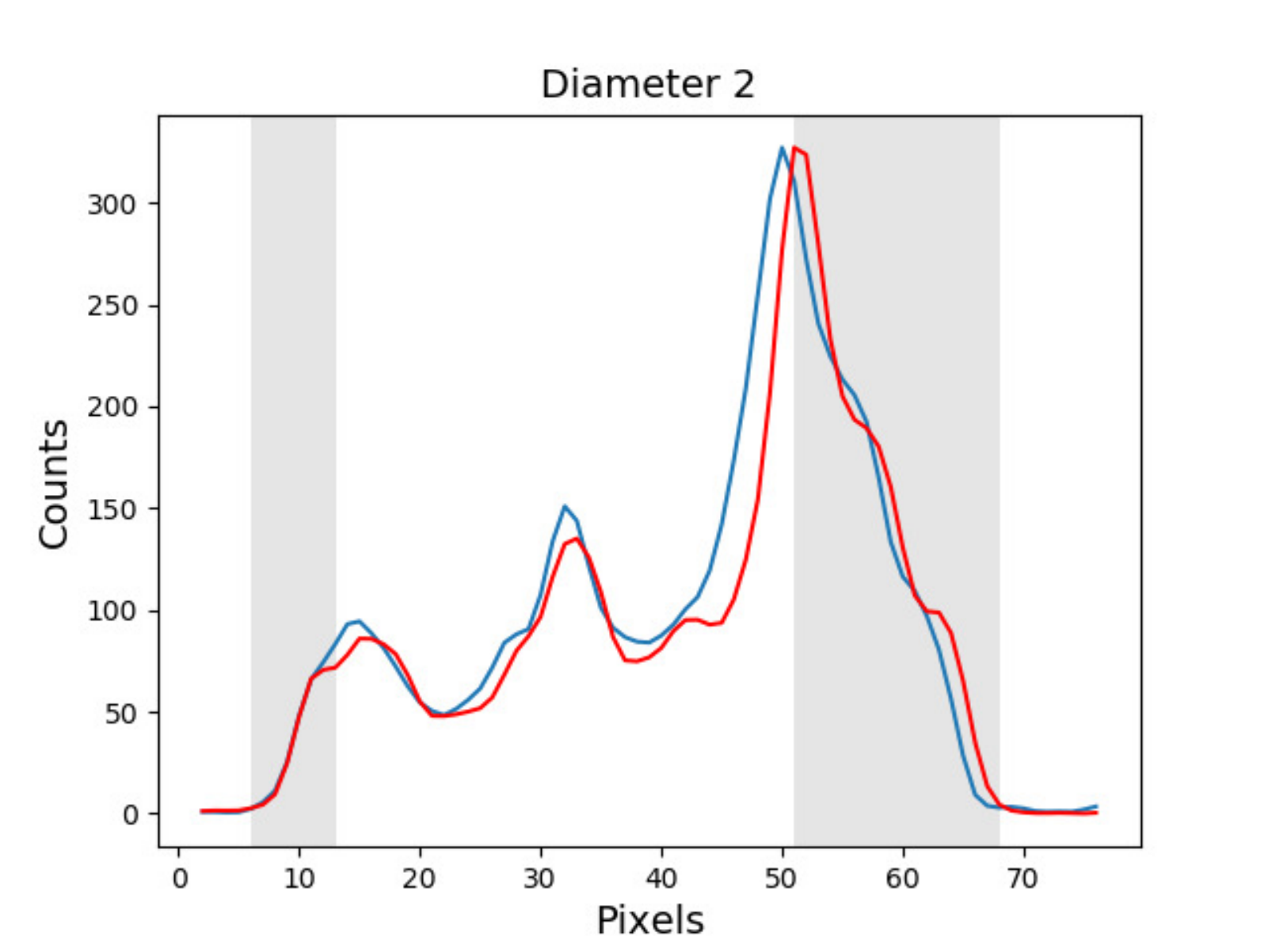}
\includegraphics[width=6cm]{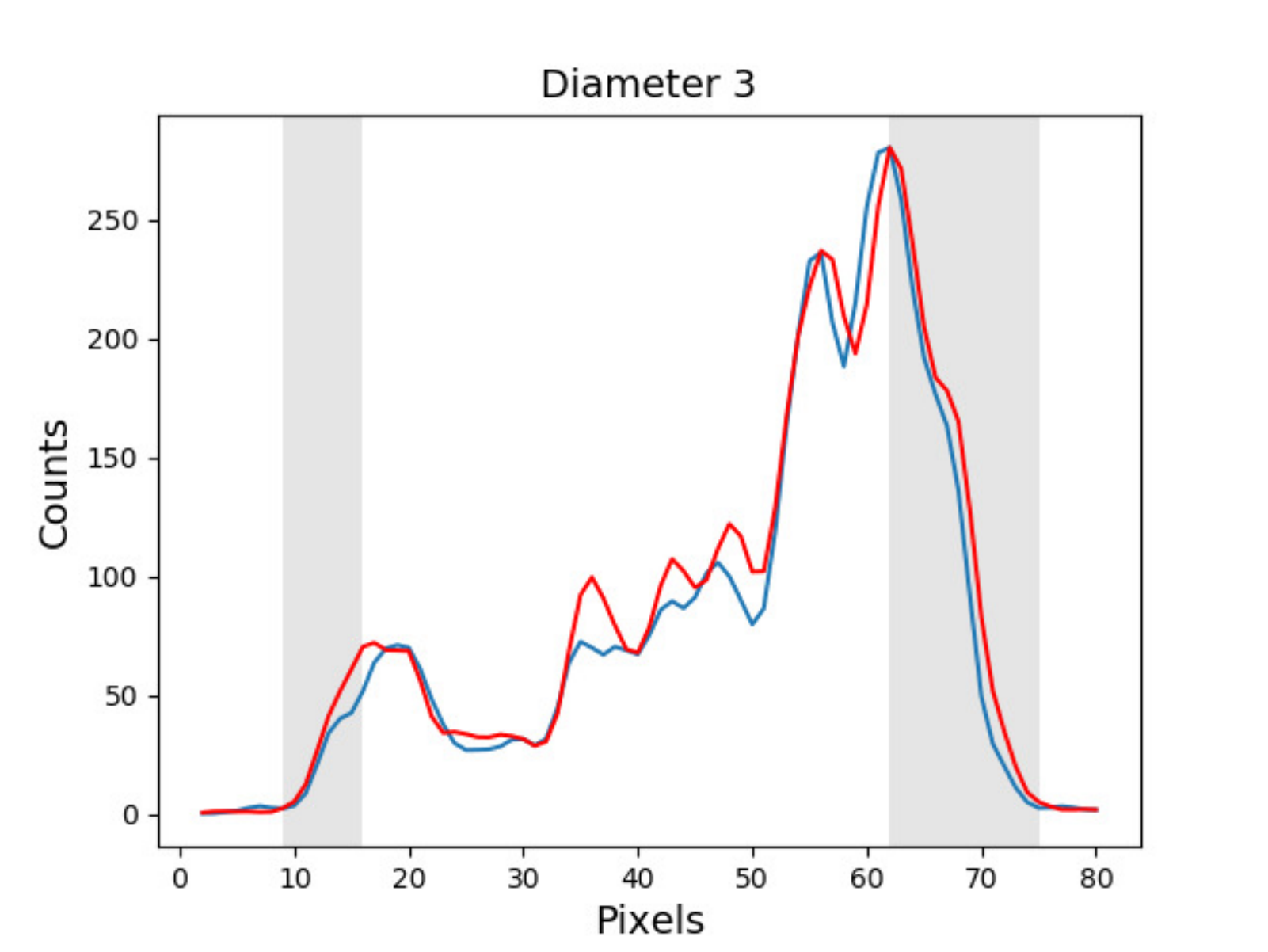}
\includegraphics[width=6cm]{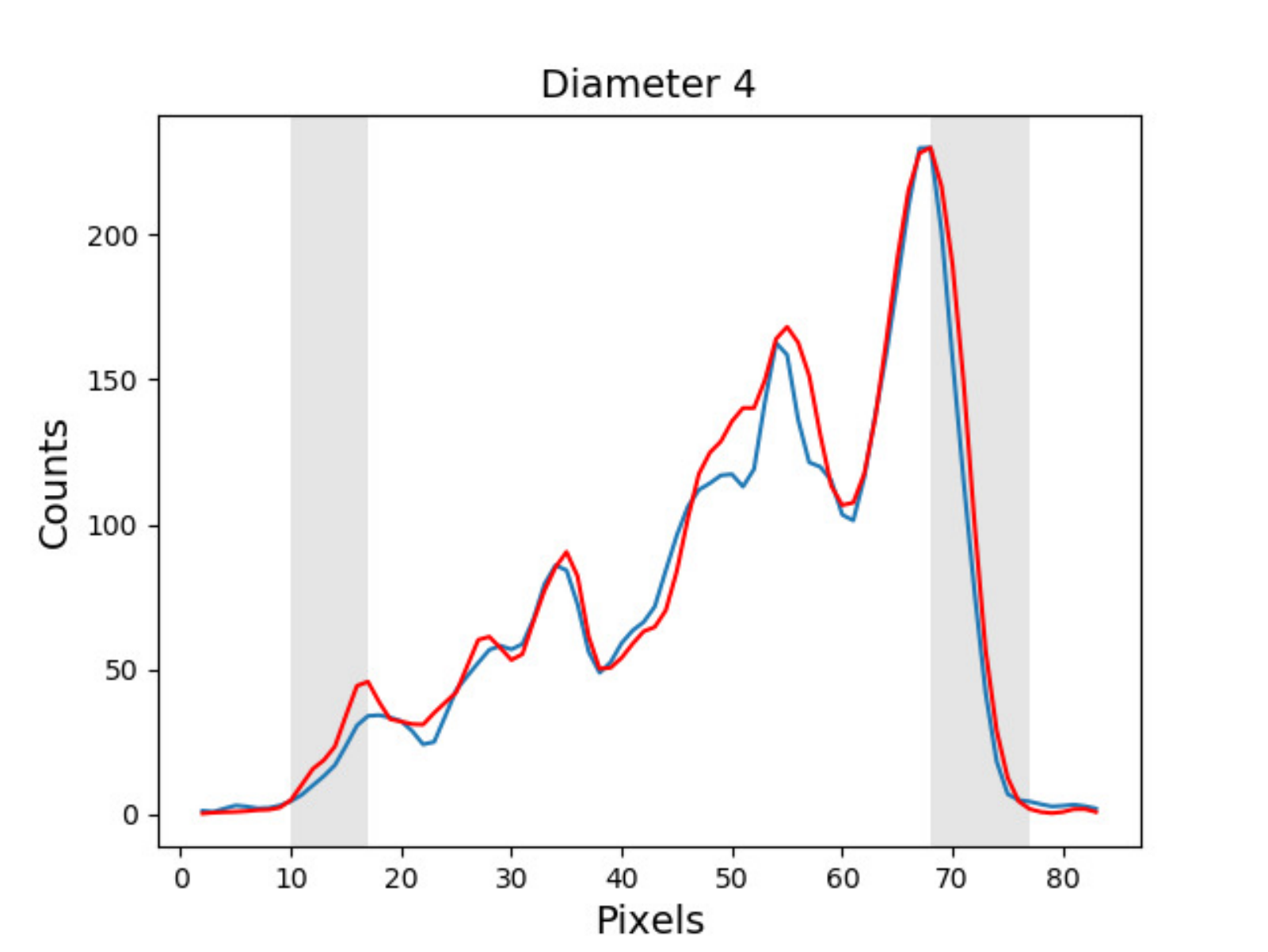}
\caption{The brightness profiles of epoch 1 (1999) in blue and epoch 2 (2017) in red for diameter regions 1 and 2, normalized for display purposes. The grey shaded regions mark the regions in which the fits were performed. The normalization was adjusted for each shaded region. The profiles run north-to-south for region 1 and east-west (left to right) for regions 2-4.
\label{diameter2}
}
\end{centering}
\end{figure}

We measured the diameter of the remnant using the radial profiles (obtained by using {\it projection} regions in {\it ds9}) along four "diameter" regions, shown in Figure~\ref{n103b}. These regions sample the entire brightness profile of the remnant along diameters covering position angles $0-180$, $45-225$, $90-270$, and $135-315$. We made no attempt to scientifically define a center of the remnant, since the site of the explosion is unknown. We simply drew the diameter regions to run through the geometric center of the circular structure of N103B. To obtain enough signal for a robust profile, each region is 10 pixels wide, or $\sim 5''$. The normalized brightness profiles extracted from each diameter in the two epochs are shown in Figure~\ref{diameter2}. We are not concerned with the small changes in internal structure, only the change in the diameter of the remnant as marked by the sharp rise of the shock front. To do this, we measure the shift in the shock front on both the left side and right side of the profiles separately (these "fit" regions are marked by shaded grey areas in Figure~\ref{diameter2}). The relative normalization of the peak of emission is tailored to each of these regions. Taking into account the uncertainty on each profile data point (not shown in the Figures for display purposes), we shift epoch 1 with respect to epoch 2 on a fine grid of $0.0048''$ resolution elements ($\sim 0.01$ {\it Chandra} pixels).

The total expansion velocity in each region is simply reported as the average of the two values we measure, one from each side of the remnant. As can be seen from the Figures, the rise in the brightness profiles marking the edge of the shock front is generally fairly consistent between epochs. A few exceptions exist, such as the left (east) sides of regions 3 and 4. Nonetheless, those were included in the fitting procedure, and simply resulted in increased uncertainties in those regions. In a few places, such as the left side of region 1, epoch 2 is interior to epoch 1, leading to a negative shock velocity, almost certainly due to a coordinate registration error, as discussed above. However, the strength of this fitting procedure is that the change of the diameter of the remnant does not depend on this. For example, in the case of diameter region 1, we "measure" a shock velocity of -4,070 km s$^{-1}$ for the left (north) side of the emission, and an incredible (and almost certainly unphysical) 14,840 km s$^{-1}$ for the right (south) side. However, the average of these two is 5,360 km s$^{-1}$, the value reported in Table~\ref{results}, which should be robust even in light of uncertainties in the image registration.

For the uncertainties, we measure both statistical and systematic error terms. The statistical uncertainties come from the fits themselves: we report the best fit as the shift in which the value of $\chi^{2}$ is minimized, and take as the 90\% confidence limits the value of the shift in each direction where $\chi^{2}$ has risen by 2.706. For the systematic uncertainties, the typical reported values, such as those in \citet{katsuda13}, of the registration uncertainties in aligning the images are irrelevant for our purposes. Instead, we found that varying the choice of fit region for each shock front, marked by the shaded grey areas, resulted in slightly different values for the shock velocity. These errors were generally small, usually a few tens of km s$^{-1}$, and are dwarfed by the statistical uncertainties on the fit. Nonetheless, we include both uncertainties, added linearly, in our results reported in Table~\ref{results}. Our values for the average expansion velocity range from 2,990 to 5,360 km s$^{-1}$, with a mean expansion velocity of 4,170 km s$^{-1}$, with lower and upper limits of 2,860 and 5,450, respectively.

\begin{figure}[htb]
\begin{centering}
\includegraphics[width=8cm]{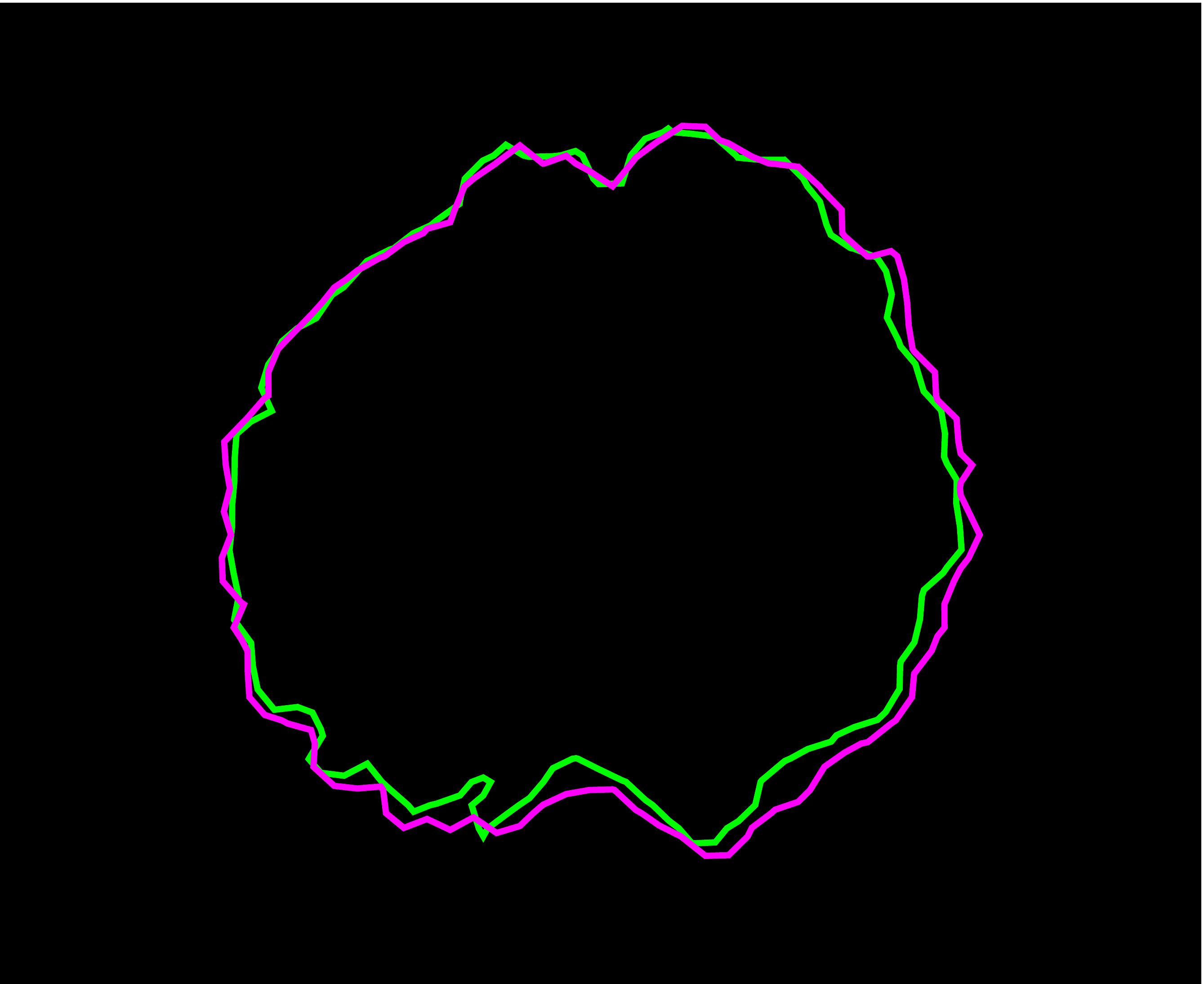}
\caption{A single contour, marking the edge of the remnant in each epoch, with epoch 1 in green, and epoch 2 in magenta. The contours are smoothed with a 2-pixel gaussian.
\label{contours}
}
\end{centering}
\end{figure}

As a final "sanity check" on the expansion of N103B, we conducted an entirely different and independent experiment. We drew a single contour in {\it ds9} around the remnant in both epochs. Since the vast majority of background pixels in a given {\it Chandra} observation have zero counts, a single contour defining the edge of the emission from the remnant (and some small amount of leakage resulting from the wings of the PSF) is quite easy to define, simply be defining a contour level of "1." The contours from both epochs are shown in Figure~\ref{contours}. We converted these contour to region files, and measured the number of pixels contained within each. In the 1999 epoch, this contour contained 2,846 pixels, while in 2017 the remnant occupied 2,983 pixels. Since the area of the remnant increases as the square of the radius, this means that, {\em on average}, the radius of the remnant has increased by 2.37\%, or about 0.36 pixels. This corresponds to a shock velocity of 4,810 km s$^{-1}$. This is somewhat higher than our average, reported above, but well within the uncertainties, confirming the expansion between the two epochs.

\begin{deluxetable}{lccc}
\tablecolumns{4} 
\tablewidth{0pc} 
\tabletypesize{\footnotesize}
\tablecaption{Expansion Velocities for N103B} \tablehead{ \colhead{Region} & Mean $v_{s}$ (km s$^{-1}$) & Lower Limit & Upper Limit}

\startdata

1 & 5360 & 4080 & 6570\\
2 & 4070 & 3140 & 4940\\
3 & 4280 & 3320 & 5250\\
4 & 2990 & 910 & 5030\\
Average & 4170 & 2860 & 5450\\

\enddata

\tablecomments{Mean $v_{s}$ is total expansion of the shock front along each diameter region, divided by two, as described in the text. Distance is assumed to be 50 kpc. The lower and upper limits include both statistical and systematic uncertainties, as described in the text. All results are reported to the nearest 10 km s$^{-1}$.}
\label{results}
\end{deluxetable}

A high shock velocity confirms N103B's status as a young SNR, as reported by \citet{rest05} and \citet{lewis03}. An expansion velocity of 4,170 km s$^{-1}$ implies an undecelerated age for N103B of 850 yr, making the real age somewhat lower. It is somewhat surprising to find such a high shock velocity, given the high densities reported in W14. The most obvious caveat here is that by only measuring the change in the remnant's diameter, we cannot know if one side of the remnant is expanding faster than the other. The high densities reported in W14 came from the western half of N103B, the same half in which \citet{ghavamian17} reported lower shock velocities from the broad H$\alpha$ component. However, these Balmer line filaments are quite faint, and it is likely that only shocks in denser regions are spectroscopically detectable through their broad component. X-ray emission above 1.7 keV is dominated by intermediate-mass elements, particularly Si and S. If these lines result from the ejecta, their velocity might be different from the blast wave velocity, complicating the results. Thus, to ensure an "apples-to-apples" comparison between the shock velocity implied by the broad H$\alpha$ width and the velocity measured by proper motion, we would need to measure the optical proper motion of the H$\alpha$ filaments used, requiring a second epoch of optical imaging. Optical proper motion measurements could also in principle provide a measure of the deceleration parameter, $\theta$, further constraining the age of the remnant.

A comparison can be drawn here between N103B and Kepler's SNR, where the forward shock speeds are found to vary by about a factor of two between the north and south rims \citet{katsuda08,vink08}. If the same velocity ratio is present here (in an east-west direction), that would lead to a shock velocity of $\sim 2800$ km s$^{-1}$ in the west (much closer to the speeds seen in \citet{ghavamian17}) and $\sim 5500$ km s$^{-1}$ in the east. Such high shock velocities in N103B may imply a nonthermal synchrotron component in the X-ray spectrum, as is seen in Kepler. In a follow-up paper on detailed X-ray spectroscopy, we will explore the evidence for this component. 

\section{Conclusions}
\label{conclusions}

We re-observed the bright LMC SNR N103B with {\it Chandra} in 2017, 17.4 years after it was first observed in 1999 with the goal of measuring the expansion of the remnant. The lack of strong detections of point sources in the field of view made absolute alignment of the two epochs impossible, but we were able to measure the change in the diameter of the remnant in four different directions, yielding an average expansion of the shock front of just under 4200 km s$^{-1}$. This further supports the view that this remnant is young. The undecelerated age is 850 yr, but since some deceleration has almost certainly occurred, the real age is younger than this, entirely consistent with the estimates from light echo studies \citep{rest05}. We encourage future monitoring of this object at all wavelengths, particularly X-ray and optical, where high resolution observations can continue to refine measurements of the expansion.

\acknowledgements

We thank the anonymous referee for providing valuable comments which improved the manuscript. Support for this work was provided by the National Aeronautics and Space Administration through Chandra Award Number G06-17064 issued by the Chandra X-ray Center, which is operated by the Smithsonian Astrophysical Observatory for and on behalf of the National Aeronautics Space Administration under contract NAS8-03060. PFW acknowledges additional support from NSF, through grant AST-1714281. IRS acknowledges support from the Australian Research Council Grant FT160100028.

\end{document}